\documentclass[a4paper,10pt,twoside]{cpc-hepnp}

\usepackage{multicol}
\usepackage{graphicx}
\usepackage{booktabs}
\usepackage{amssymb,bm,mathrsfs,bbm,amscd}
\usepackage[tbtags]{amsmath}
\usepackage{lastpage}

\begin{document}

\fancyhead[co]{\footnotesize H.C. Kim et al: Hadrons from a hard
wall AdS/QCD model}


\title{Hadrons from a hard wall AdS/QCD model}

\author{%
      Hyun-Chul Kim$^{1;1)}$\email{hchkim@inha.ac.kr}%
\quad Youngman Kim$^{2;2)}$\email{ykim@apctp.org}%
\quad U.T. Yakhshiev$^{1,3;3)}$\email{yakhshiev@inha.ac.kr}}

\maketitle

\address{%
1~(Department of Physics, Inha University, Incheon 402-751, Korea)\\
2~(Asia Pasific Center for Theoretical Physics and Department of
Physics, Pohang University of Science and Technology,\\ Pohang,
Gyeongbuk 790-784, Korea)\\
3~(Department of Nuclear and Theoretical Physics, National
University of Uzbekistan, Tashkent-174, Uzbekistan) }

\begin{abstract}
We review a recent work on masses of mesons and nucleons within a hard
wall model of holographic QCD in a unified approach.  In order to
treat meson and nucleon properties on the same footing, we introduced
the same infrared (IR) cut for both sectors. In framework of present 
approach, the best fit to the experimental data is achieved
introducing the anomalous dimension for baryons.
\end{abstract}



\begin{multicols}{2}

After the realization that a string theory in anti-de Sitter (AdS)
space corresponds to the strongly coupled conformal field theory
(CFT) in its 
boundary\cite{Maldacena:1997re,Gubser:1998bc,Witten:1998qj}, several
models inspired by this AdS/CFT correspondence have been developed to
investigate hadron properties in the low-energy region. 
In general, these models can be divided into two categories: In
the first approach, so-called "top-down" approach, one starts from the 
fundamental (string/M) theory living on AdS$_{d+1}\times {\cal C}$
(where ${\cal C}$ is a compact manifold) and tries to formulate an 
effective theory which is supposed to describe low energy phenomena
in strongly interacting systems. In the second approach,
so-called "bottom-up" approach, one starts from the 5 dimensional (5D) 
phenomenological Lagrangian, which incorporates as much as possible
properties of the fundamental theory (QCD) and known low-energy
phenomenological facts, and attempts to extend the model by
constructing its dual (which include the extra dimensions too). For
the detailed descriptions of mesons in gauge/gravity duality in both 
approaches, refer to a recent review(for example, see
ref.~\cite{Erdmenger:2007cm}). 

It is well known that QCD is not a conformal theory. Consequently, the
different approaches have been developed in the "bottom-up" approach,
where one has some free parameters in a constructed effective
model. Clearly, these free parameters are 
usually related to phenomenologically well known quantities. One of
such parameters is the size of the extra dimension (compactification
scale), which is fixed and related to the QCD scale  
$\Lambda_{\rm QCD}$. Naturally, the compactification breaks the
conformal invariance and these types of models are coined as {\em
  hard-wall models}\cite{Erlich:2005qh,DaRold:2005zs}. 

We start from a hard-wall model for mesons developed in
refs.\cite{Erlich:2005qh,DaRold:2005zs} and study its applications
to nucleons\cite{Hong:2006ta,HCKim09}. The model has a geometry of
5D AdS
\begin{equation}
ds^2 \;=\; g_{MN} dx^M dx^N \;=\; \frac1{z^2}(\eta_{\mu\nu}dx^\mu
dx^\nu-dz^2)\,,
\end{equation}
where $\eta_{\mu\nu}$ stands for the 4D Minkowski metric:
$\eta_{\mu\nu} = \mathrm{diag}(1,-1,-1,-1)$. The 5D AdS space is
compactified by two different boundary conditions, i.e. the IR
boundary at $z=z_m$ and the UV one at $z=\epsilon\to 0$. Considering
the global chiral symmetry $\mathrm{SU(2)}_L\times\mathrm{SU(2)}_R$
of QCD, we need to introduce 5D local gauge fields $A_L$ and $A_R$
of which the values at $z=0$ play a role of external sources for
$\mathrm{SU(2)}_L$ and $\mathrm{SU(2)}_R$ currents respectively.
Since chiral symmetry is known to be broken to $\mathrm{SU(2)}_V$
spontaneously as well as explicitly, we introduce a bi-fundamental
field $X$ with respect to the local gauge symmetry
$\mathrm{SU(2)}_L\times\mathrm{SU(2)}_R$, in order to realize the
spontaneous and explicit breakings of chiral symmetry in the AdS
side. Considering these two, we can construct the bi-fundamental 5D
bulk scalar field $X$ in terms of the current quark mass $m_q$ and
the quark condensate $\sigma$
\begin{equation}
X_0(z) \;=\; \langle X \rangle \;=\; \frac12(m_q z + \sigma z^3)
\end{equation}
with isospin symmetry assumed.

The 5D gauge action in AdS space with the scalar bulk field and the
vector field  can be expressed as
\begin{eqnarray}
S_M \!\!\!\!&=&\!\!\!\!\int d^4x\int dz\,  {\cal L}_M\,,\nonumber\\
{\cal L}_M\!\!\!\!&=&\!\!\!\!\frac1{z^5}\, \mathrm{Tr} \left[ |DX|^2
+3|X|^2 -\frac{1}{2g_5^2}(F_L^2 + F_R^2) \right],
\label{eq:eff_meson}
\end{eqnarray}
where covariant derivative and field strength tensors are defined as
$DX \;=\;
\partial X - i A_L X + iXA_R$ and $F_{L,R}^{MN}=\partial^M
A_{L,R}-\partial^N A_{L,R}-i[A^M_{L,R},\,A^N_{L,R}] $.  The 5D gauge
coupling $g_5$  is fixed by matching the 5D vector correlation
function to that from the operator product expansion (OPE):
$g_5^2=12\pi^2/N_c$.  The 5D mass of the bulk gauge field $A_{L,R}$
is determined by the relation $m_5^2=(\Delta - p)(\Delta + p
-4)$\cite{Gubser:1998bc,Witten:1998qj} where $\Delta$ denotes the
canonical dimension of the corresponding operator with spin $p$ and
turns out to be $m_5^2=0$ because of gauge symmetry. The effective
action~(\ref{eq:eff_meson}) describes the mesonic
sector\cite{Erlich:2005qh,DaRold:2005zs} completely apart from
exotic mesons\cite{Kim:2008qh}.

To consider baryons in the flavor-two ($N_F=2$) sector, one needs to
introduce a bulk Dirac field corresponding to the nucleon at the
boundary\cite{Hong:2006ta,HCKim09}. This hard-wall model was 
applied to describe the neutron electric dipole
moment\cite{Hong:2007tf} and holographic nuclear
matter\cite{Kim:2007xi}.  In this model, the nucleons are the
massless chiral isospin doublets $(p_L,\, n_L)$ and $(p_R,\,n_R)$
which satisfy the 't Hooft anomaly matching. Then the spontaneous
breakdown of chiral symmetry induces a chirally symmetric mass term
for nucleons
\begin{equation}
\mathcal{L}_{\chi SB} \;\sim\; -M_N \left( \begin{array}{c} \bar{p}_L \\
    \bar{n}_L  \end{array}\right)\, \Sigma\, (p_R, \,n_R) + \mathrm{h.c.},
\end{equation}
where $\Sigma=\exp(2i\pi^a\tau^a/f_\pi)$ is the nonlinear
pseudo-Goldstone boson field that transforms as $\Sigma\to U_L
\Sigma U_R^\dagger$ under $\mathrm{SU(2)}_L\times\mathrm{SU(2)}_R$.
The $\tau^a$ and $f_\pi$ represent the SU(2) Pauli matrices and the
pion decay constant, respectively. Thus, we have to consider the
following mass term in the AdS side
\begin{equation}
\mathcal{L}_{\mathrm{I}} \;=\; -g \left( \begin{array}{c} \bar{p}_L \\
    \bar{n}_L  \end{array}\right)\, X\, (p_R, \,n_R) + \mathrm{h.c.},
\end{equation}
where $g$ denotes the mass coupling (or Yukawa coupling) between $X$
and nucleon fields, which is usually fitted by reproducing the
nucleon mass $M_N=940$ MeV. In this regard, we can introduce two 5D
Dirac spinors $N_1$ and $N_2$ of which the Kaluza-Klein (KK) modes
should include the excitations of the massless chiral nucleons
$(p_L,\, n_L)$ and $(p_R,\,n_R)$, respectively.  By this
requirement, one can fix the IR boundary conditions for $N_1$ and
$N_2$ at $z=z_m$.

Note that the 5D spinors $N_{1,2}$ do not have chirality. However,
one can resolve this problem in such a way that the 4D chirality is
encoded in the sign of the 5D Dirac mass term. For a positive 5D
mass, only the right-handed component of the 5D spnior remains near
the UV boundary $z\to 0$, which plays the role of a source for the
left-handed chiral operator in 4 dimension. It is vice versa for a
negative 5D mass.  The 5D mass for the $(d+1)$ bulk dimensional
spinor is determined by the AdS/CFT expression
\begin{equation}
(m_5)^2 \;=\; \left(\Delta -\frac{d}{2}\right)^2 \label{eq:5dmass}
\end{equation}
and turns out to be $m_5=5/2$. However, since QCD does not have
conformal symmetry in the low-energy regime, the 5D mass might
acquire an anomalous dimension due to a 5D renormalization flow.
Though it is not known how to derive it, we will introduce some
anomalous dimension to see its effects on the spectrum of the
nucleon.

Summarizing all these facts, we are led to the 5D gauge action for
the nucleons
\begin{eqnarray}
S_N &=& \int d^4x\int dz\, \frac1{z^5}\, \mathrm{Tr} \left[{\cal
    L}_{\mathrm{K}}+{\cal L}_{\mathrm{I}} \right], \cr
{\cal L}_{\mathrm{K}} &=&  i \bar{N}_1
 \Gamma^M \nabla_M N_1 + i \bar{N}_2
 \Gamma^M \nabla_M N_2 \nonumber\\
 &&- \frac52 \bar{N}_1 N_2 +  \frac52 \bar{N}_2 N_1
\nonumber\\
{\cal L}_{\mathrm{I}} &=& -g\left[ \bar{N}_1 X N_2 + \bar{N}_2
  X^\dag N_1\right],
\label{EffAct}
\end{eqnarray}
where
\begin{equation}
 \nabla_M \;=\; \partial_M + \frac{i}{4}\, \omega_M^{AB} \Gamma_{AB}
 -i A_M^L\,.
\end{equation}
The non-vanishing components of the spin connection are
$w_M^{5A}=-w_M^{A5}=\delta_M^A/z$ and
$\Gamma_{AB}=\frac1{2i}[\Gamma^A,\Gamma^B]$ are the Lorentz
generators for spinors.  The $\Gamma$ matrices are related to the
ordinary $\gamma$ matrices as $\Gamma^M= (\gamma^\mu,\,-i\gamma_5)$.

The details of the calculations within the present approach can be
found in ref.\cite{HCKim09}.
Presenting our results we note, that the most of input parameters of
the model such as $m_q$, $\sigma$ and $z_m$ are can be well fitted
from the data in the mesonic sector\cite{Erlich:2005qh}. In order to
reproduce the data in the baryonic sector we have only one free 
parameter $g$. Our calculations showed that this additional
parameter alone, which is bounded to some region $|g|<g_{\rm
crit}$\cite{HCKim09}, is not enough to reproduce the experimental
data.
\begin{center}
\includegraphics[width=7cm]{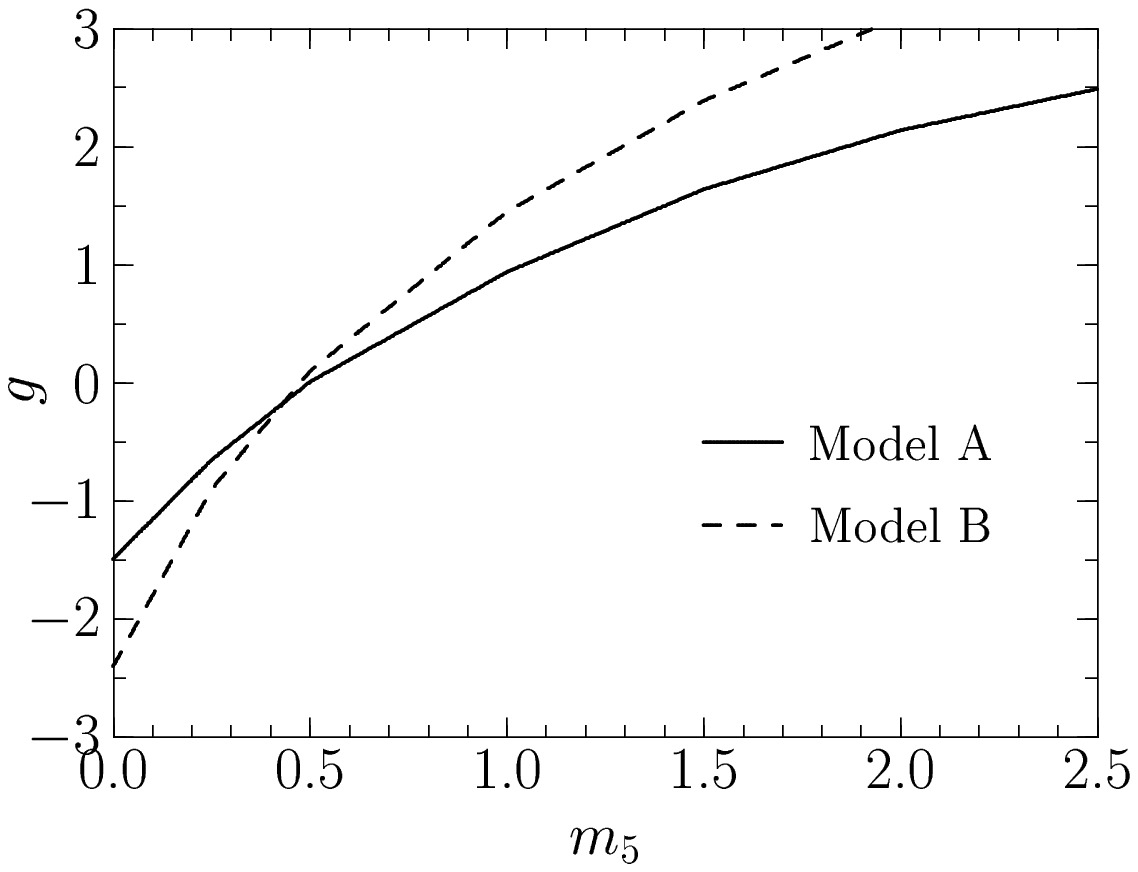}
\figcaption{\label{fig1}  Mass coupling $g$ dependence on the 
renormalized 5D mass $m_5$. The parameters for models A (solid curve)
and B (dashed one) are 
taken from ref.\protect\cite{Erlich:2005qh}.}
\end{center}
\begin{center}
\includegraphics[width=7cm]{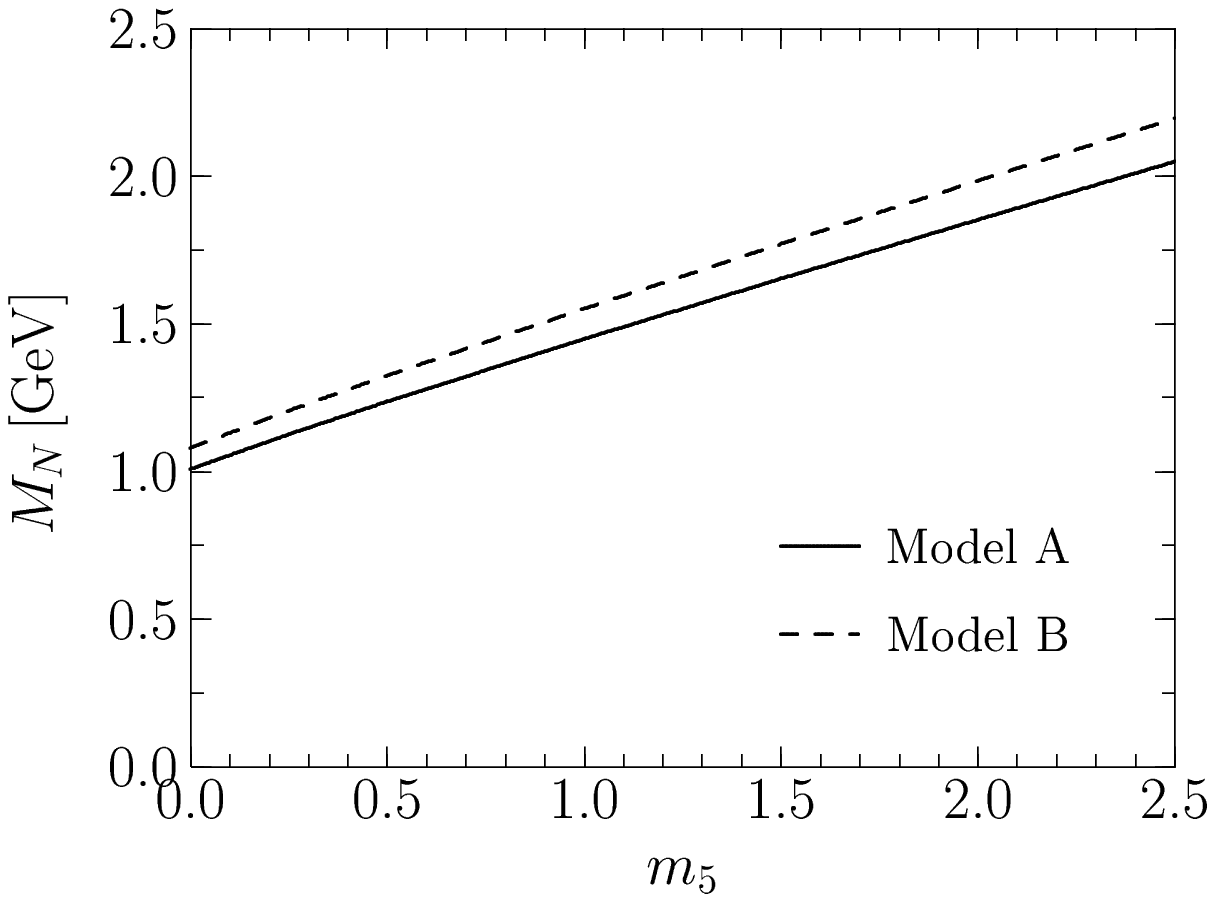}
\figcaption{\label{fig2} The mass of the lowest-lying nucleon (in 
GeV) as a function of the renormalized 5D mass $m_5$. The 
notations are same as in fig.\ref{fig1}. }
\end{center}

For example, with the given sets of input parameter sets A and B taken
from the mesonic sector\cite{Erlich:2005qh}, the reproduced baryon
masses are to 
high: low lying nucleon state mass is $m_N=2050$~MeV (Model A) and
$m_N=2196$~MeV (Model B) in contrast to the experimental value
940~MeV. One has to look for additional possibilities in order to
reproduce experimental results and one of such possibilities is
to introduce the anomalous dimension for baryons as we mentioned
above. Although it is not obvious to calculate an anomalous
dimension, it is known that it renormalizes the 5D mass. Taking this 
assumption into account, one can consider $m_5$ as a free parameter.

The results are drawn in figs.~\ref{fig1} and~\ref{fig2}.
One can see that, when anomalous dimension is set to zero (i.e.
$m_5=5/2$ is fixed), the experimental data is badly reproduced. The 
inclusion of an additional parameter (i.e. considering $m_5$ as a
free parameter) improves the output data well but leads to 
larger values of the possible anomalous dimension.

The best fit to experimental data in the chiral limit is reproduced with 
the values of input parameters $z_m^{-1}=285$~MeV,
$\sigma^{1/3}=227$~MeV, and $g=-9.6$,  as listed in table~\ref{tab1}. 
\begin{center}
\tabcaption{ \label{tab1}  The best fit to the experimental data. 5D
mass is equal to zero due to the large anomalous dimension.}
\footnotesize
\begin{tabular*}{80mm}{c@{\extracolsep{\fill}}lccccc}
\toprule &$(p,n)^+$ & $N^+$ &$N^-$&$\rho$&$\rho$\\
\hline Experiment&\quad939&1440&1535&776&1475\\
This model&\quad930 & 1826 & 1856&685&1573\\
\bottomrule
\end{tabular*}
\end{center}
One can note that the masses of nonstrange baryons are well
reproduced, within (10$\sim$20)\,\% deviations from the experimental
data.

\acknowledgments{The work of U.~Yakhshiev is partially supported by
AvH foundation. The present work is also supported by Basic Science
Research Program through the National Research Foundation of Korea
(NRF) funded by the Ministry of Education, Science and Technology
(grant number: 2009-0073101). Y.K. acknowledges the Max Planck
Society(MPG) and the Korea Ministry of Education, Science and
Technology(MEST) for the support of the Independent Junior Research
Group at the Asia Pacific Center for Theoretical Physics (APCTP). }

\end{multicols}

\vspace{-2mm}
\centerline{\rule{80mm}{0.1pt}}
\vspace{2mm}

\begin{multicols}{2}

\end{multicols}

\clearpage

\end{document}